\documentclass[12pt]{iopart}
\usepackage{graphicx}
\usepackage{dcolumn}
\usepackage{bm}
\usepackage[tight]{subfigure}
\usepackage{iopams}
\usepackage{verbatim}
\usepackage{color}

\begin{document}

\title{Thermoelectric signatures of a Majorana bound state coupled to a quantum dot}

\author{Martin Leijnse}

\address{
  Solid State Physics and Nanometer Structure Consortium (nmC@LU),
  Lund University, 221 00 Lund, Sweden 
}

\begin{abstract}
We theoretically investigate the possibility to use thermolectric measurements to detect Majorana bound states and to investigate their coupling 
to a dissipative environment. The particle-hole symmetry of Majorana states would normally lead to a vanishing Seebeck coefficient, 
i.e., a vanishing open-circuit voltage resulting from a temperature gradient. We discuss how coupling to a quantum dot with a gate-controlled 
energy level breaks particle-hole symmetry in a 
tunable manner. The resulting gate-dependent Seebeck coefficient provides a new way to evidence the existence of Majorana states, which can be combined 
with conventional tunnel spectroscopy in the same setup. Furthermore, the thermoelectric properties rely on the ability of the quantum dot-Majorana 
system to sense the temperature of the bulk superconductor and can be used to extract information about the dissipative decay of 
Majorana states, which is crucial for quantum information applications.

\end{abstract}
\pacs{
  74.25.fg, 
  85.35.Gv, 
  74.45.+c, 
  74.78.Na, 
}
\maketitle
\section{Introduction}
During the last few years there has been a remarkable interest in the search for quasiparticle excitations which mimic the properties of 
Majorana fermions~\cite{Wilczek09, Alicea12rev, Beenakker11rev, Leijnse12rev}. Majorana-like quasiparticles were first predicted to occur in the 
$\nu = 5/2$ fractional quantum Hall state~\cite{Moore91}, but it was more recently realized that they could be engineered by 
proximity-inducing superconductivity in materials with strong spin-orbit coupling, such as the surface states (or edge states) of a topological 
insulator~\cite{Fu08, Nilsson08, Fu09, Linder10}, or a two-dimensional semiconductor quantum well~\cite{Sau10, Alicea10, Akhmerov11}.
Currently, much experimental interest is devoted to one-dimensional semiconductor nanowires~\cite{Oreg10, Lutchyn10} 
with strong spin-orbit coupling, such as InSb or InAs. When brought into proximity with a superconductor (SC) and exposed to a magnetic field,
such wires can become effectively spinless $p$-wave SCs, shown by Kitaev~\cite{Kitaev01} to host Majorana bound states (MBS) at 
the end points. 

Tunnel spectroscopy can be used to verify the existence of a MBS, for example at the end of a nanowire, since it gives rise to a characteristic 
conductance peak at zero bias voltage~\cite{Bolech07, Law09, Flensberg10}. Several recent 
experiments~\cite{Mourik12, Deng12, Das12, Churchill13, Finck13} have indeed observed such a peak in the expected parameter regimes,  
indicating the existence of MBS.
However, other studies have shown that similar conductance peaks may arise also from e.g., Kondo physics~\cite{DeFranceschi13, Chang13}, weak 
anti-localization~\cite{Pikulin12}, and subgap states~\cite{Potter10, Kells12, Lee13}, making MBS identification only through tunnel spectroscopy 
somewhat problematic.
Alternatively, signatures of MBS could be found e.g., in the quantized conductance steps in a more transparent junction 
(quantum point contact)~\cite{Wimmer11}, or in Josephson junctions~\cite{Kitaev01, Jiang11, Williams12, Rokhinson12}, but all these experiments 
are associated with significant difficulties and it is not clear whether they can provide a true unique fingerprint of MBS.

Thermoelectric measurements can provide more information about a systems electronic properties than can be inferred from the conductance. Here, the central quantity is the Seebeck coefficient, $S = -V_\mathrm{th} / \Delta T$, defined by the relation between an applied temperature difference $\Delta T$ and the resulting open-circuit voltage $V_\mathrm{th}$. $S$ is related to the energy-derivative of the conductance and can therefore be used to distinguish between electron dominated ($S < 0$) and hole-dominated ($S > 0$) electric transport. A MBS has perfect electron-hole symmetry and should therefore not contribute to a non-zero $S$ (even when coupling between MBS give rise to a finite-energy state). This is in contrast to sharp conductance resonances arising from, e.g, a highly peaked density of states (DOS)~\cite{Mahan96, Humphrey05}, resonant levels~\cite{Beenakker92, Staring93, Svensson12} (which give $S=0$ only when perfectly aligned with the Fermi level), or the Kondo effect~\cite{Scheibner05, Costi10}. Combined measurements of $G$ and $S$ could therefore provide further evidence of MBS. However, also normal Andreev bound states could give zero contribution to $S$ and, conversely, a Majorana junction could have a non-zero $S$ if particle-hole symmetry is broken by some other mechanism, such as potential barriers or accidentally formed quantum dots (QDs) close to the tunnel probe. A non-zero $S$ was indeed predicted in Ref.~\cite{Hou13} for a one-dimensional Majorana mode, if it is  coupled to a normal electrode with energy-asymmetric density of states. We also note that thermoelectric measurements have been 
proposed as a way to probe non-Abelian statistics in quantum hall systems~\cite{Yang09}.

\begin{figure}[t!]
  \includegraphics[height=0.5\linewidth]{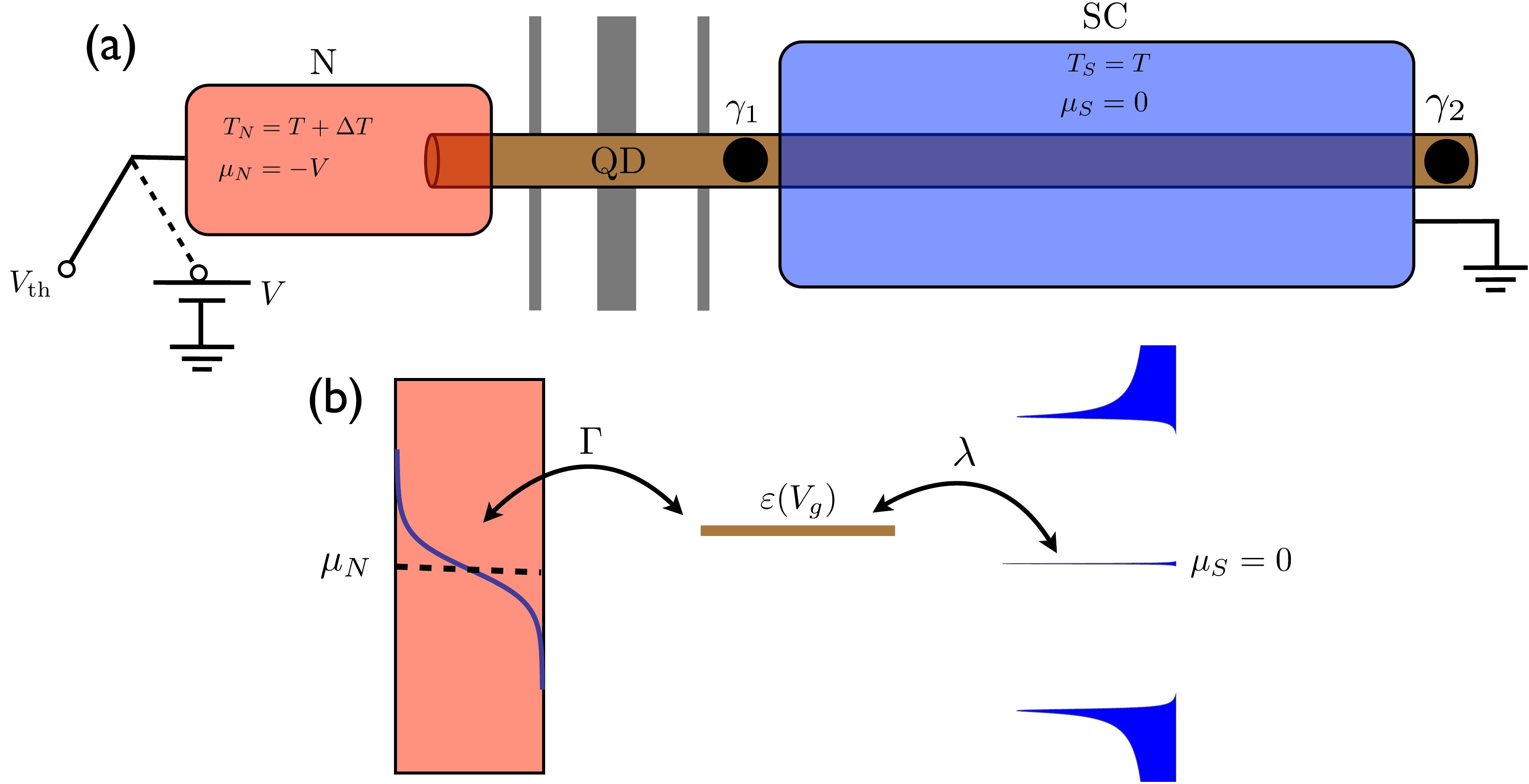}	
	\caption{\label{fig:setup}
	(a) Sketch of N-QD-MBS junction. The SC is grounded and kept at the ambient temperature. When measuring the conductance $G$ 
	(thermoconductance $G_T$), N is electrically (thermally) biased. When measuring $S$, N is thermally biased and the open-circuit (thermal) voltage 
	$V_\mathrm{th}$ is measured. The gray lines represent narrow gates which define the QD and control $\varepsilon$. $\gamma_{1,2}$ 
	represent the two MBS.
	(b) Level diagram of the setup in (a), showing the thermal smearing of the Fermi surface in N (curved line) and indicating tunneling between 
	N and the QD with rate $\Gamma$ and between the QD and MBS with amplitude $\lambda$.
	}
\end{figure}
Here, we wish to investigate a MBS transport junction where the thermoelectric properties are experimentally tunable, possibly providing stronger evidence of MBS.
We focus on a system where a QD is intentionally formed between the normal metal tunnel probe (N) and the MBS (N-QD-MBS junction), see sketch in figure~\ref{fig:setup}. This setup allows for a controlled breaking of particle-hole symmetry by shifting the QD energy level with a gate voltage $V_g$. Previous studies~\cite{Leijnse11, Zitko11, Golub11, Liu11, Lee12} have shown that the conductance of similar systems can reveal the existence of MBS.
We show that $S(V_g)$, which can be measured in the same setup as the conductance, can provide complementary evidence of MBS. The N-QD-MBS system does, however, not guarantee a non-zero $S$, which also requires the combined QD-MBS system to acquire a different temperature than that which is provided by the Fermi sea in N. $S$ therefore depends on the coupling between the combined QD-MBS system and the dissipative environment provided by the bulk SC, the temperature of which differs by $\Delta T$ from that of N. We discuss how this allows thermoelectric measurements to provide information about the nature of the dissipative coupling and reveal, for example, relaxation of the fermion parity quantum number, even without coupling between different MBS. Such information is crucial for the use of MBS in topological quantum computation schemes~\cite{Nayak08rev}.

The paper is organized as follows. In section~\ref{sec:model} we introduce the minimal N-QD-MBS model used in the paper. Section~\ref{sec:seebeck} introduces 
a simple theory (and analytic results) for the thermoelectric properties, assuming that the QD-MBS system acquires the same temperature as the bulk superconductor.
In section~\ref{sec:env} we relax the condition of a perfectly thermalized QD-MBS system and instead consider two different models for the 
interactions with the dissipative environment formed by the bulk superconductor, which give rise to different thermoelectric responses.
Finally, section~\ref{sec:conclusions} summarizes and concludes.

\section{N-QD-MBS junction}\label{sec:model}
Figure~\ref{fig:setup} shows a sketch of the N-QD-MBS junction. A normal metallic electrode is tunnel coupled to a quantum dot, which in turn is tunnel coupled to the edge of a SC (figure~\ref{fig:setup} shows an example setup with a semiconducting nanowire, but the discussion is not limited to this specific geometry). We consider the simplest possible model representing this setup, described by the Hamiltonian $H = H_N + H_D + H_S + H_T^{ND} + H_T^{DS}$, where 
\begin{eqnarray}
\label{eq:H_N}
	H_N 	&=&	\sum_{k} \left( \varepsilon_k - \mu_N \right) n_k, \\
\label{eq:H_D}
	H_D	&=&	\varepsilon n_D, \\
\label{eq:H_M}
	H_M	&=& 	\frac{i}{2} \xi \gamma_1 \gamma_2 = \xi \left( n_f - \frac{1}{2} \right), \\
\label{eq:H_ND}
	H_T^{ND} &=& 	t \sum_k d^\dagger c_k + h.c., \\
\label{eq:H_DS}
	H_T^{DS} &=&    \left( \lambda d - \lambda^{*} d^\dagger \right) \gamma_1 =  \lambda d f^\dagger + \lambda d f + h. c..
\end{eqnarray}
Here, $n_k = c_k^\dagger c_k$ is the number operator for non-interacting electrons in state $k$ in N. N is biased with chemical potential $\mu_N = -V$ (we use units where $e = k_B = \hbar = 1$) and kept at temperature $T_N = T + \Delta T$.
$n_D = d^\dagger d$ describes the quantum dot with a single level controlled by a gate voltage, $\epsilon = -\alpha_g V_g$, where $\alpha_g$ is the gate coupling. We consider the case where a large magnetic field has been applied to induce a topological superconducting phase, thereby also introducing a large Zeemann splitting on the dot. We therefore only consider one spin species of dot electrons and can then neglect also the opposite spin in N (the transport effects of spin and Coulomb blockade was investigated in Ref.~\cite{Leijnse11}).  
The Hamiltonian~(\ref{eq:H_M}) is a low-energy description of the edge of the SC in the topological regime~\cite{Kitaev01}, which includes only the two MBS (localized on opposite ends of the wire) and therefore is valid for energies well within the superconducting gap. $\gamma_i$ are MBS operators fulfilling $\gamma_i^\dagger = \gamma_i$ and $\{ \gamma_i, \gamma_j \} = 2 \delta_{ij}$. $\xi$ is the coupling between the two MBS, which vanishes exponentially with their separation. The second form of Eq.~(\ref{eq:H_M}) is written instead in terms of the operator $f = (\gamma_1 + i \gamma_2)/2$, which describes a standard (but nonlocal) fermionic state with occupation $n_f = f^\dagger f$. We assume the SC to be grounded with chemical potential $\mu_S = 0$. In addition, the bulk SC is kept at a fixed temperature $T_S = T$, but we defer the discussion on how this temperature affects the low-energy MBS to section~\ref{sec:env}.

$H_T^{ND}$ describes the coupling between N and the QD with amplitude $t$. We neglect the $k$-dependence of both $t$ and the normal electrode DOS $\rho$, which leads to an energy-independent tunnel coupling $\Gamma = 2 \pi \rho |t|^2$, setting the inverse time scale for single-electron tunneling between N and the QD. Tunneling between the QD and the SC is described by $H_T^{DS}$, where we have again projected the SC Hamiltonian close to midgap onto the MBS represented by $\gamma_1$ (we assume negligible coupling between the QD and $\gamma_2$). Written in terms of the fermion operator $f$, the tunneling is seen to contain anomalous terms ($\propto d f, f^\dagger d^\dagger$) which do not conserve particle number.
This reflects the fact that the $n_f$ state is an equal superposition of an electron and a hole, and can be filled ($n_f = 0 \rightarrow n_f = 1$) either by adding
an electron ($d f^\dagger$) or a hole ($f^\dagger d^\dagger$) to the SC from the QD. 

\section{Gate-dependent Seebeck coefficient}\label{sec:seebeck}
\begin{figure}[t!]
  \includegraphics[height=0.3\linewidth]{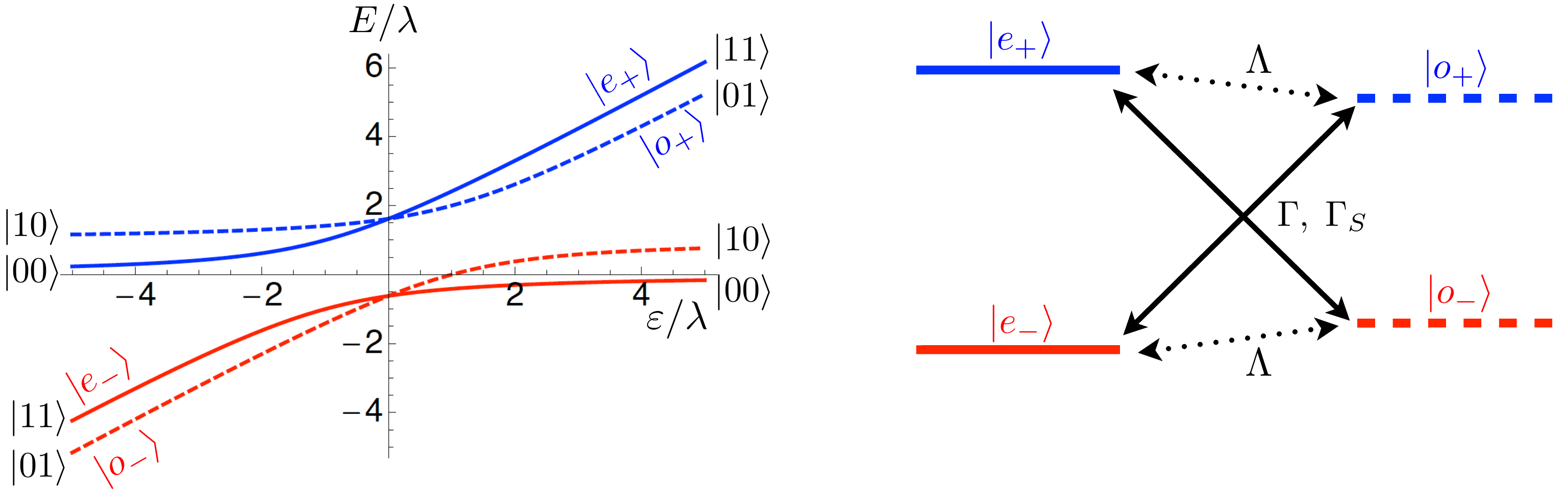}	
	\caption{\label{fig:eigen}
	Left: Eigenenergies $E$ of $H_0$ as a function of $\varepsilon$ with $\xi = \lambda$. For $| \varepsilon / \lambda| \gg 1$, 
	the eigenstates are approximately equal to the indicated number states. Right: Level diagram showing the coupling of eigenstates via coupling to 
	a dissipative environment. Tunneling to N ($\Gamma$) or to subgap states in the SC ($\Gamma_S$) changes $n_D$
	and therefore couples $|e_\pm\rangle$ and $|o_\mp\rangle$ most strongly (solid lines). Quasiparticle 
	poisoning ($\Lambda$) changes $n_f$, thus primarily coupling $|e_\pm\rangle$ and $|o_\pm\rangle$ (dotted lines).}
\end{figure}

We now want to calculate the electric current $I$ flowing out of N as a result of an applied electric bias $\mu_N = -V$ and temperature difference 
$T_N = T + \Delta T$. We want to use N as a weakly coupled probe of the QD-MBS system and therefore consider the limit $\Gamma \ll T, \lambda$. 
In this regime, it is most instructive to start by exactly diagonalizing the QD-MBS system, described by $H_0 = H_D + H_M + H_T^{DS}$. 
We use the basis $|n_f n_D\rangle = (f^\dagger)^{n_f}(d^\dagger)^{n_D}|00\rangle$ of eigenstates of $n_f$ and $n_D$. As mentioned above, $H_T^{DS}$ 
does not conserve particle number, but $H_0$ conserves instead the parity of the total fermion number in the QD and the SC 
(i.e., $n_D + n_f$ being even or odd) and is thus block diagonal with an even block $H_0^e$, acting on $\{|00\rangle, |11\rangle\}$, 
and an odd block $H_0^o$, acting on $\{ |10\rangle$, $|01\rangle\}$
\begin{eqnarray}
\label{eq:He}
	H_0^e 	&=& \left( \begin{array}{cc} 0 & \lambda \\ \lambda^* & \varepsilon + \xi \end{array} \right), \\
\label{eq:Ho}
	H_0^o 	&=& \left( \begin{array}{cc} \xi & \lambda \\ \lambda^* & \varepsilon \end{array} \right). 
\end{eqnarray}
The eigenstates are 
$|e_{+}\rangle = \alpha_{e} |00\rangle + \beta_{e} |11\rangle$,
$|e_{-}\rangle = \beta_{e} |00\rangle - \alpha_{e} |11\rangle$,
$|o_{+}\rangle = \alpha_{o} |10\rangle + \beta_{o} |01\rangle$,
$|o_{-}\rangle = \beta_{o} |10\rangle - \alpha_{o} |01\rangle$,
shown as a function of $\varepsilon$ in figure~\ref{fig:eigen}. 
For $| \varepsilon / \lambda| \gg 1$, the eigenstates are close to pure number states, but the QD-MBS coupling mixes the number states 
when $| \varepsilon / \lambda| \lesssim 1$ and leads to avoided crossings as a function of $\varepsilon$.

We now include the effect of tunneling between N and the QD described by $H_T^{ND}$. 
An exact treatment is possible, but in the limit $\Gamma \ll \lambda, T$ considered here, it is sufficient and far more
intuitive to treat $H_T^{ND}$ within lowest non-vanishing order perturbation theory.  Then the current, in terms of the occupation 
probabilities $P_a$ of the eigenstates $|a\rangle = |e_\pm\rangle,  |o_\pm \rangle$ of $H_0$, is given by
\begin{eqnarray}
\label{eq:I}
	I 		&=& \sum_{a a'} W^I_{a a'} P_{a'}.	
\end{eqnarray}
The current rate matrix
\begin{eqnarray}
\label{eq:WI}
	W^I_{a a'}  	&=& \Gamma_{a a'} f \left( \frac{E_{a a'} - \mu_N} {T_N} \right)
			- \Gamma_{a' a} \left[ 1 - f \left( \frac{E_{a'a} - \mu_N}{T_N} \right) \right] 	
\end{eqnarray}
describes the rate for a process connecting eigenstates $|a\rangle$ and $|a'\rangle$ by removing (first term) or adding (second term) an electron 
from/to N (the same eigenstates are coupled by both these processes because $H_0$ does not conserve particle number). 
Here $f(x) = 1 / (e^x + 1)$ is the Fermi function, $E_{a a'} = E_a - E_{a'}$ is the difference between eigenenergies, and 
$\Gamma_{a a'} = \Gamma (|\langle a |d| a'\rangle |^2 + |\langle a |d^\dagger| a'\rangle |^2$), which is only non-zero if $a$ and $a'$ have different 
fermion parities.
Since tunneling changes the dot occupation it mainly couples $|e_\pm\rangle$ and $|o_\mp\rangle$, 
which for $|\varepsilon / \lambda| \gg 1$ correspond to number states with different $n_D$, 
see figure~\ref{fig:eigen} [the relaxation processes ($\Gamma_S$ and $\Lambda$) in the level diagram will be 
discussed in section~\ref{sec:env}]. However, close to the anti-crossing, the eigenstates are equal mixtures of the number states and 
tunneling couples also $|e_\pm\rangle$ and $|o_\pm\rangle$.

\begin{figure}[t!]
  \includegraphics[height=0.17\linewidth]{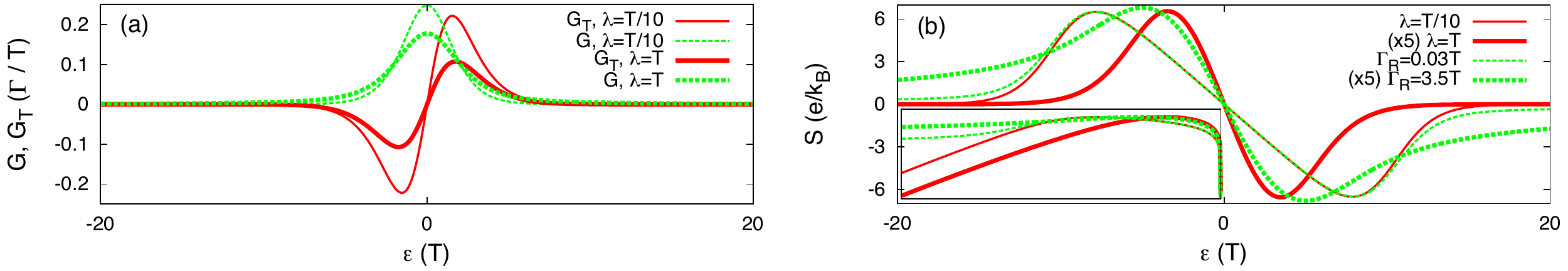}	
	\caption{\label{fig:simpleresults}
	(a) $G_T$ (red solid lines) and $G$ (green dashed lines) as a function of $\varepsilon$, controlled by $V_g$, with $\lambda = T/10$ (thin lines) 
	and $\lambda = T$ (thick lines). (b) $S$ as a function of $\varepsilon$ for the N-QD-MBS junction (red solid lines) with 
	$\lambda = T/10$ (thin lines) and $\lambda = T$ (thick lines), compared with a N-QD-N junction with $\Gamma_R \gg \Gamma_L$ and 
	$\Gamma_R$ adapted to give approximately the same peak height as the results with a MBS. The inset shows the same plot for $\varepsilon < 0$, 
	but on logarithmic scale. $\xi = 0$ was assumed in all plots. 
	Note that the curves with large $\lambda$ and $\Gamma_R$ have been multiplied by 5.}
\end{figure}

In an experiment, $S = -V_\mathrm{th} / \Delta T$ is measured in an open-circuit configuration from the voltage which builds up as a result of the applied 
temperature difference. 
From a theory perspective, it is more convenient to consider the closed circuit with a finite current. In linear response, meaning small 
$V$ and $\Delta T$, the current is given by $I = G V + G_T \Delta T$, from which the Seebeck coefficient is found to be $S = -G_T / G$ by
setting $I = 0$. The advantage of the open-circuit experiment is that even when $G$ and $G_T$ are both very small, $S$ can be large and perhaps easier 
to measure.
We focus first on the thermoconductance $G_T$, i.e., the current response to a small temperature bias $\Delta T$ with $V = 0$. 
The result is especially simple for $\xi = 0$, where the even and odd parity sectors are equivalent (we can then use a simplified notation 
where the equivalent states $e_{+}$ and $o_+$ are both denoted $+$, while $e_{-}$ and $o_-$ are denoted $-$).
Inserting the eigenstates and eigenenergies of Eqs.~(\ref{eq:He}) and~(\ref{eq:Ho}) into Eqs.~(\ref{eq:I}) and~(\ref{eq:WI}) we find
\begin{eqnarray}
\label{eq:GTsimple}
	G_T  		&=& \frac{2 \Gamma \left( |\alpha|^4 - |\beta|^4 \right)}{\Delta T}   
			\left\{ \left[ 1 - f\left( \frac{E_{+-}}{T + \Delta T} \right) \right] P_+ 
			- f\left( \frac{E_{+-}}{T + \Delta T} \right) P_-\right\}. \nonumber \\
\end{eqnarray}
The temperature of the QD-MBS system enters through the ratio between excited state and 
ground state occupations, $P_+ / P_-$. If we assume this ratio to follow the Boltzmann distribution with the 
temperature $T_S = T$ of the bulk SC (a non-trivial assumption as we will see below), Eq.~(\ref{eq:GTsimple}) becomes
\begin{eqnarray}
\label{eq:GTsimple2}
	G_T  		&=& \frac{\Gamma \left( |\alpha|^4 - |\beta|^4 \right)}{\Delta T} 
			    \left[  f \left( \frac{E_{+-}}{ T + \Delta T} \right) - f \left( \frac{E_{+-}}{ T } \right)\right] \nonumber \\ \\ 
\label{eq:GTsimplelinear}
			&=&	 \frac{\Gamma \left( |\alpha|^4 - |\beta|^4 \right) E_{+-}} {T^2} f' \left( \frac{E_{+-}}{T} \right) + O\left( \Delta T \right). 
\end{eqnarray}
$G_T (\varepsilon)$ is plotted in figure~\ref{fig:simpleresults}(a). 
At the point where the dot level crosses the Fermi energy, $\varepsilon = 0$, particle-hole symmetry is restored and
$G_T = 0$ since $|\alpha| = |\beta|$. Away from this point $G_T$ increases linearly, but then decays again since $E_{+-} \approx |\varepsilon|$
for $\varepsilon \gg \lambda$ and thus $f '\left( E_{+-} / T \right) \rightarrow 0$ as $T / \varepsilon \rightarrow 0$.  

The conductance is found to be 
\begin{eqnarray}
\label{eq:Gsimple}
	G		&=& \frac{\Gamma}{V} \left\{ |\alpha|^4 \left[ f \left( \frac{- E_{+-} + V}{T} \right) - 
						f \left( \frac{- E_{+-}}{T} \right) \right] \right. \nonumber \\
			&+& \left. |\beta|^4  \left[ f \left( \frac{ E_{+-} + V}{T} \right) - f \left( \frac{ E_{+-}}{T} \right) \right] \right. \nonumber \\
			&-& \left. |\alpha|^2|\beta|^2 \left[ f \left( \frac{V}{T} \right) - f \left( \frac{-V}{T} \right)\right] \right\} \\
\label{eq:Gsimplelinear}
			&=& \frac{\Gamma}{T} \left[ \left(|\alpha|^4 + |\beta|^4 \right) f'\left( \frac{ E_{+-}}{T} \right) - \frac{1}{2} |\alpha|^2 |\beta|^2 \right] 
				+ O\left( V \right), 
\end{eqnarray}
which is plotted in figure~\ref{fig:simpleresults}(a). 
Here we have assumed that, since $\Gamma \ll \lambda$, the bias drops only at the N-QD junction, such that $\varepsilon$ is independent of $V$.
The Seebeck coefficient is
\begin{eqnarray}
\label{eq:Ssimple}
	S		&=& \frac{|\beta|^4 - |\alpha|^4}{\left(|\alpha|^4 + |\beta|^4\right) - \frac{1}{2}|\alpha|^2|\beta|^2 / f'\left( E_{+-} / T\right)} \; \frac{E_{+-}}{T},
\end{eqnarray}
which is plotted as a function of $\varepsilon$ in figure~\ref{fig:simpleresults}(b). 
It is instructive to compare our result to the well-known expression for sequential tunneling in a single-level QD coupled to two 
normal electrodes~\cite{Beenakker92}, $S = - \varepsilon / T$. 
In fact, Eq.~(\ref{eq:Ssimple}) resembles this result since $E_{+-} \approx \varepsilon$ for $|\varepsilon / \lambda| \gg 1$, but Eq.~(\ref{eq:Ssimple}) 
approaches zero for large enough $\varepsilon$. Most strikingly, the derivative of the Fermi function in the second term in the denominator gives rise 
to an exponential decay for $|\varepsilon / T| \gg 1$.
%
The linear regime is, however, limited also in a QD coupled to two normal electrodes when taking cotunneling or broadening 
of the QD level into account~\cite{Turek02, Svensson12}. Therefore, in figure~\ref{fig:simpleresults}(b) we compare our result from Eq.~(\ref{eq:Ssimple}) 
with $S(\varepsilon)$ of a QD coupled to two normal electrodes (L and R), i.e., a N-QD-N junction, calculated within a scattering 
formalism~\cite{Landauer57, Buttiker86}. We choose the tunnel couplings 
asymmetric, $\Gamma_L \ll \Gamma_R$, to resemble the N-QD-MBS setup with $\Gamma \ll \lambda$. $S$ is then independent of $\Gamma_L$ and we adjust 
$\Gamma_R$ to obtain approximately the same peak values of $S$ as for the case with a MBS. Although the lineshapes are similar close to resonance, 
the decay is clearly different, being algebraic for a N-QD-N junction but exponential for the N-QD-MBS junction because of the term in Eq.~(\ref{eq:Ssimple}) 
containing the derivative of the Fermi function. The difference is clearer when plotting the result on a logarithmic scale, see inset in 
figure~\ref{fig:simpleresults}(b).
From a measurement of $S(\varepsilon)$ it is thus possible to distinguish between a QD coupled to a MBS and one coupled to a normal electrode 
(where the "normal electrode" could be the finite subgap DOS in a SC with a soft gap).
Sharp conductance peaks due to the Kondo effect would also give a different $S(\varepsilon)$~\cite{Scheibner05, Costi10}. 
Re-introducing normal units, the peak values of $S$ in figure~\ref{fig:simpleresults}(b) corresponds to a thermal voltage of more than 500 
$\mu \mathrm{V}/\mathrm{K}$ (for $\lambda = T/10$), which can readily be measured even with a small $\Delta T$ on the order of tens of mK.

\section{Coupling to the environment}\label{sec:env}
In section~\ref{sec:seebeck} we assumed the QD-MBS system to be in thermal equilibrium at a temperature $T$ which differs from that in N, 
but we did not specify how thermal equilibrium is reached.
In general, within our perturbative approach we should instead find the occupation probabilities $P_a$ of the eigenstates of $H_0$ from 
rate equations
\begin{eqnarray}
\label{eq:ME}
	\dot{P}_a 	&=& 0 = \sum_{a'} \left( W_{a a'} P_{a'} - W_{a' a} P_{a}\right), \\
\label{eq:Pnorm}
	1		&=& \sum_{a} P_a.
\end{eqnarray}
Equation~(\ref{eq:ME}) expresses that the change in occupation of $|a\rangle$, which is zero in the steady state, is given by the sum of all 
ingoing processes, minus the sum of all outgoing processes, each weighted by the occupation of the corresponding initial state. 
Equation~(\ref{eq:Pnorm}) enforces probability normalization. 
If we find the rate matrix $W_{a a'}$ from lowest order perturbation theory 
in $H_T^{ND}$, it is the same as $W_{a a'}^I$ in Eq.~(\ref{eq:WI}), except that the second term comes with a plus sign. 
To find $G_T$ we solve Eqs.~(\ref{eq:ME}) and~(\ref{eq:Pnorm}) with $V = 0$. This gives $P_a / P_a' = \mathrm{exp}(E_{a a'}/T_N)$, i.e., Boltzmann 
distributed according to the temperature of N, as imposed by the Fermi distribution of the tunneling electrons. 
This is a direct consequence of the lack of dissipation in the SC within the low-energy Hamiltonian~(\ref{eq:H_M}):
The MBS can carry a current due to the lack of charge conservation, but has zero width and does not allow energy to be dissipated and
can therefore not act as a thermal bath.
Inserting this distribution of $P_a$ in Eq.~(\ref{eq:GTsimple}) (or the corresponding expression with $\xi > 0$), we find $G_T = 0$ 
and therefore also $S = 0$, as expected since no real temperature difference exists. 

\begin{figure*}[t!]
  \includegraphics[height=0.33\linewidth]{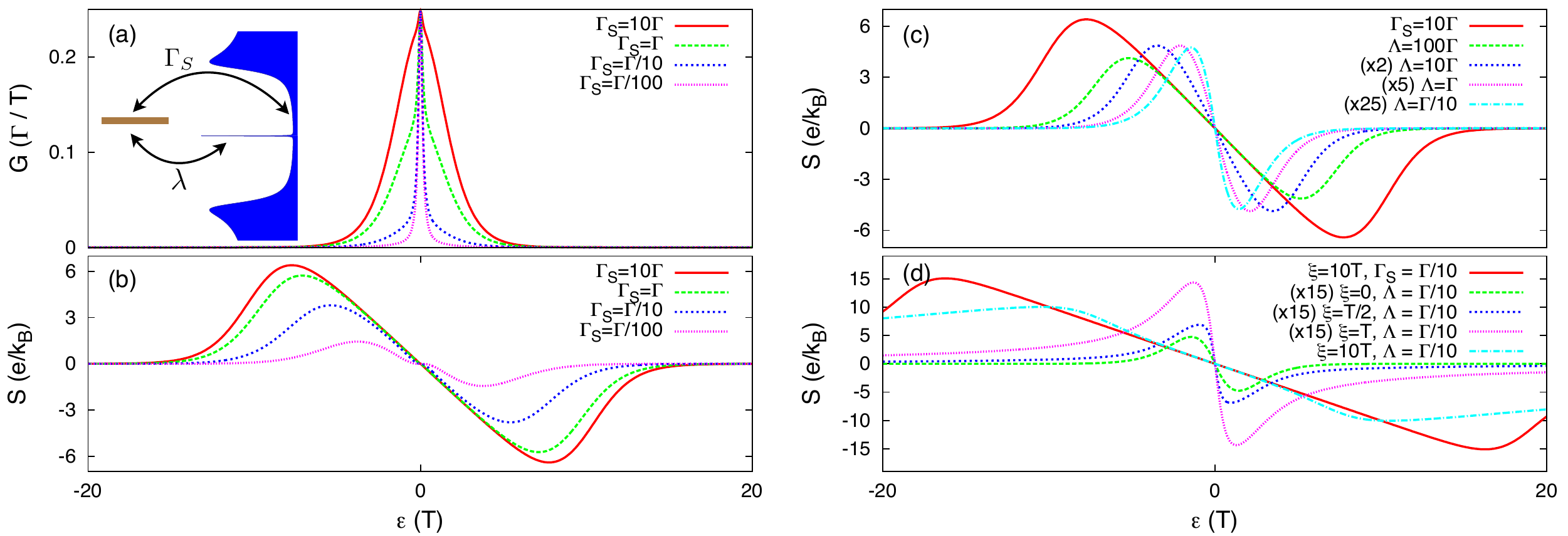}	
	\caption{\label{fig:relax}
	(a) $G(\varepsilon)$ with $\lambda = T/10$, $\xi = 0$, and different values of $\Gamma_S / \Gamma$. 
	The inset shows a quantum dot level coupled to a MBS and to subgap states in a soft-gap SC. 
	(b) $S(\varepsilon)$ for the same parameters as in (a). 
	(c) Comparison of $S(\varepsilon)$ with coupling to subgap states and with quasiparticle poisoning, with all other parameters as in (a) and (b). 
	Note that the curves with $\Lambda = 10 \Gamma, \Gamma, \Gamma/10$ have been multiplied by $2, 5, 25$.
	(d) $S(\varepsilon)$ with $\lambda = T/10$, $\Lambda = \Gamma / 10$, $\Gamma_S = 0$, and increasing coupling $\xi$ between the MBS. 
	For comparison, the result with $\Lambda = 0$, $\Gamma_S = \Gamma / 10$ is also shown for $\xi = 10 T$.}
\end{figure*}

To have a finite thermoelectric effect and motivate the results found in section~\ref{sec:seebeck} we need to couple $H_0$ to an additional 
dissipative environment, held at the temperature of the SC. One such source of dissipation is a finite continuous 
DOS inside the superconducting gap~\cite{Takei13, Stanescu13}, often observed in experiments where superconductivity is proximity-induced in e.g., 
a nanowire~\cite{Mourik12, Deng12, Das12, Churchill13, Finck13}, see inset of figure~\ref{fig:relax}(a). 
We model the continuum of subgap states as an additional normal electrode with a Hamiltonian analogous to 
$H_N$ and a coupling to the QD analogous to $H_T^{ND}$. Assuming the subgap DOS to be constant within the relevant energy window of 
width $\sim \mathrm{max} (\lambda, T, |\varepsilon|)$, this gives rise to equivalent contributions to $W_{a a'}$ as the coupling to N, but proportional 
to a different tunnel coupling $\Gamma_S$ and with the Fermi functions evaluated at $T_S = T$, $\mu_S = 0$. There is no direct contribution to 
the current rate matrix $W_{aa'}^I$ since we evaluate the current in N. 
Such a perturbative treatment neglects processes $\propto \Gamma \Gamma_S$ and is valid when 
$\Gamma, \Gamma_S \ll \lambda$. Close to resonance, i.e., when $|\varepsilon / \lambda| \lesssim 1$, transport between the QD and the SC is then dominated 
by tunneling into the MBS, but the subgap states can impose a temperature different from $T_N = T + \Delta T$ on the QD-MBS system. 

Figure~\ref{fig:relax}(a) shows $G(\varepsilon)$ and figure~\ref{fig:relax}(b) shows $S(\varepsilon)$ 
for $\xi = 0$, $\lambda = T / 10$ and for different values of $\Gamma_S / \Gamma$ 
(for $\Gamma, \Gamma_S \ll \lambda$, $S$ only depends on this ratio, not directly on $\Gamma$ or $\Gamma_S$).
For $\Gamma_S \gg \Gamma$, the results are identical to what was found in section~\ref{sec:seebeck}.  
When $\Gamma_S \lesssim \Gamma$, heating N also heats the QD-MBS system to a temperature $T + \Delta T \;\Gamma / (\Gamma + \Gamma_S)$
and $G_T$ is reduced since the effective temperature difference becomes smaller. The width of the peak in $G(\varepsilon)$ is 
$\sim \lambda$ for $\Gamma_S \ll \Gamma$, but $\sim T$ for $\Gamma_S \gg \Gamma$ when tunneling from the QD to the subgap states in 
the SC gives a substantial contribution (the peak height is independent of $\Gamma_S$
since $\lambda \gg \Gamma_S$ guarantees that tunneling into the MBS dominates at $|\varepsilon / \lambda| \lesssim 1$).
When $\Gamma_S \lesssim \Gamma$, the peak shows a crossover between different slopes, basically being a sum of a high peak with a 
narrow width $\sim \lambda$ and a low peak with a larger width $\sim T$. There is also a corresponding crossover in $S(\varepsilon)$, between a small  
slope in the $\lambda$-dominated regime and a larger slope in the $T$-dominated regime, most clearly seen in the magenta curve in figure~\ref{fig:relax}(b). 
When $\lambda \gtrsim T$ (not shown), the conductance is independent of $\Gamma_S$, the width of the peak in $G(\varepsilon)$ is $\sim \lambda$ 
and the only effect of reducing $\Gamma_S/ \Gamma$ is a smaller $G_T$ and therefore smaller $S$.

In addition to the above considered relaxation due to tunneling into subgap states, which changes the QD occupation $n_D$, 
we also investigate relaxation due to quasiparticles tunneling into one of the edge MBS, thereby changing the occupation number $n_f$. 
Such quasiparticle poisoning is a known problem in superconducting charge qubits~\cite{Mannik04, Aumentado04} and has been considered 
also for topological SC with MBS~\cite{Leijnse11, Budich12, Rainis12}. We do not specify the source of the quasiparticles or a microscopic model, 
but employ a simple phenomenological description with thermally distributed 
quasiparticles (at temperature $T_S = T$) with constant density of states.
The relaxation rates which should be added to $W_{a a'}$ are then
\begin{eqnarray}
	\Lambda_{a a'}		&=& \Lambda \sum_{i = 1,2}| \langle a |\gamma_i |a'\rangle|^2 
				\times \left\{ \begin{array}{l} 
					1 \; \; \mathrm{if} \; E_{a'} > E_{a} \\
					e^{E_{a' a}/T} \; \; \mathrm{otherwise} \end{array} \right. 
\end{eqnarray}
$S(\varepsilon)$ for different coupling strengths $\Lambda$ are shown in figure~\ref{fig:relax}(c). 
For very large $\Lambda$, the results 
are similar to figures~\ref{fig:simpleresults}(b) and~\ref{fig:relax}(b), but for smaller $\Lambda$, $S$ becomes significantly suppressed and 
the lineshape is much more narrow. The reason for the difference in relaxation due to $\Gamma_S$ and due to $\Lambda$ lies in the eigenstates 
which are coupled by the corresponding processes, see sketch in figure~\ref{fig:eigen}. $\Gamma_S$, like $\Gamma$, changes $n_D$ and therefore mainly couples 
$|e_\pm\rangle$ and $|o_\mp\rangle$, which always have different energies. $\Lambda$ instead changes $n_f$ and thus primarily couples 
$|e_\pm\rangle$ and $|o_\pm\rangle$, which for $\xi = 0$ have exactly the same energy. Two uncoupled MBS form a zero-energy fermionic state, 
the occupation of which is always $\langle n_f \rangle = 1/2$ in thermal equilibrium, regardless of the temperature. A dissipative coupling of 
$|e_\pm\rangle$ and $|o_\pm\rangle$ can therefore not impose a temperature on the QD-MBS system at $\xi = 0$, which  
adopts the temperature of N leading to $G_T = 0$ and $S=0$.

The finite $G_T$ results from the eigenstates being superpositions of different number states, which gives a 
small coupling $\tilde{\Lambda}(\varepsilon) = \Lambda \alpha^2(\varepsilon) \beta^2(\varepsilon)$ 
also of $|e_\pm\rangle$ and $|o_\mp\rangle$. The effective temperature of the QD-MBS system is then given by
$T + \Delta T \;\Gamma / (\Gamma + \tilde{\Lambda}(\varepsilon))$. $G_T$, and therefore $S$, vanish when this effective temperature 
approaches $T_N = T + \Delta T$, i.e., when $\alpha^2 \beta^2 \ll \Gamma / \Lambda$. 
The narrow lineshape of $S(\varepsilon)$ results from $\alpha^2 \beta^2 \rightarrow 0$ when $\lambda / \varepsilon \rightarrow 0$.
However, also $G$ is affected by the quasiparticle poisoning and the precise form of $S(\varepsilon)$ is nontrivial. 
Nonetheless, if the subgap states can be suppressed such that $\Gamma_S \approx 0$, a measurement of $S(\varepsilon)$ 
allows for an estimate of $\Lambda$, either an order-of-magnitude estimate based on the decrease and changed lineshape of 
$S(\varepsilon)$ for $\Lambda \sim \Gamma$, or a more accurate estimate based on comparison with numerical results.
Note that $\Gamma$ can be found from a conductance measurement and can be controlled with gates in a setup as in 
figure~\ref{fig:setup}.

The above discussion suggests that when $\Lambda$ dominates relaxation, $S$ should be very sensitive to a coupling between the two MBS at 
the opposite ends of the wire, resulting in a finite energy $\xi$ associated with occupation of the fermionic $n_f$ state.
This is indeed seen in figure~\ref{fig:relax}(d), where both the magnitude and lineshape of $S(\varepsilon)$ is drastically
affected by an increasing $\xi$. 
For comparison, the result is shown for the largest value of $\xi$ also for $\Gamma_S$-dominated relaxation.
In general, whenever there is significant relaxation due to subgap states like in figure~\ref{fig:relax}(b), or when thermal 
equilibrium is assumed like in figure~\ref{fig:simpleresults}(b), the peak value of $S$ increases with increasing $\xi$. However,
the effect is qualitatively different and much smaller than for $\Lambda$-dominated relaxation:
There is no effect unless $\xi \gtrsim T$ and the slope remains the same in the range of $\varepsilon$ where $S(\varepsilon)$ 
grows linearly, but this linear range is increased (both $G$ and $G_T$ are, however, individually suppressed by $\xi > 0$). 


\section{Conclusions}\label{sec:conclusions}
In this work, we have theoretically investigated the thermoelectric properties of a N-QD-MBS junction. The QD level breaks particle-hole symmetry, 
which is otherwise perfect for a MBS, and leads to a finite Seebeck effect. The Seebeck coefficient is calculated as a function of the QD level 
position and is shown to be different for the QD-MBS system than for other origins of sharp conductance resonances. Thermoelectric measurements can 
therefore provide evidence of MBS complementary to standard conductance measurements, where both experiments can be done in the same setup. 
A further advantage of thermoelectric measurements is that, unlike the conductance, the Seebeck coefficient remains large even for very weak 
tunnel coupling to the MBS.  
In addition, we have shown that the Seebeck coefficient is sensitive to the nature and strength of the coupling of the QD-MBS system to 
its dissipative environment. For example, the result is markedly different when dissipation is dominated by subgap states in the SC compared to 
when it is dominated by quasiparticle poisoning. If quasiparticle poisoning dominates dissipation, the associated rate could be estimated from the 
shape and amplitude of $S(\varepsilon)$, even without coupling between the two end MBS.

\ack
I am grateful to Heiner Linke, Hongqi Xu, and Karsten Flensberg for discussions and feedback on the 
manuscript.
Financial support from the Swedish Research Council (VR) is gratefully acknowledged.

\end{document}